\documentclass[conference]{IEEEtran}
\usepackage[utf8]{inputenc}
\usepackage[english]{babel}
\usepackage[T1]{fontenc}
\usepackage{graphicx}
\usepackage{colortbl}
\usepackage{xspace}

\usepackage{booktabs}
\usepackage{multirow}

\usepackage{pgfplots}
\pgfplotsset{compat=1.9}

\usepackage{algorithm}
\usepackage{algpseudocode}
\usepackage{amssymb}
\usepackage{amsmath}
\usepackage{fancyhdr}
\usepackage{varwidth}

\usepackage{balance}

\title{Comparing Constraints Mined From Execution Logs to Understand Software Evolution}
\author{
\IEEEauthorblockN{
Thomas Krismayer\IEEEauthorrefmark{1}\IEEEauthorrefmark{2}, 
Michael Vierhauser\IEEEauthorrefmark{2}, 
Rick Rabiser\IEEEauthorrefmark{1}\IEEEauthorrefmark{2}, 
Paul Gr\"{u}nbacher\IEEEauthorrefmark{1}\IEEEauthorrefmark{2}}
\IEEEauthorblockA{\IEEEauthorrefmark{1}Christian Doppler Laboratory MEVSS, \IEEEauthorrefmark{2}Institute for Software Systems Engineering\\ Johannes Kepler University Linz, Austria\\
{\{thomas.krismayer, michael.vierhauser, rick.rabiser, paul.gruenbacher\}}@jku.at
}
}

\begin{document}
\renewcommand{\headrulewidth}{0pt}
\pagestyle{fancy}

\maketitle
\fancyhead[C]{Accepted for publication at the  2019  IEEE Int'l Conf. on Software Maintenance and Evolution \\ Final Version available at: https://doi.org/10.1109/ICSME.2019.00082}

\definecolor{darkBlue}{rgb}{0.2, 0.2, 0.6}
\newcommand{\keyword}[1]{\textcolor{darkBlue}{\textbf{#1}}}

\newcommand{\code}[1]{\small\textsf{#1}\normalsize}

\begin{abstract}
Complex software systems evolve frequently, e.g., when introducing new features or fixing bugs during maintenance. However, understanding the impact of such changes on system behavior is often difficult. Many approaches have thus been proposed that analyze systems before and after changes, e.g., by comparing source code, model-based representations, or system execution logs. In this paper, we propose an approach for comparing run-time constraints, synthesized by a constraint mining algorithm, based on execution logs recorded before and after changes. Specifically, automatically mined constraints define the expected timing and order of recurring events and the values of data elements attached to events. Our approach presents the differences of the mined constraints to users, thereby providing a higher-level view on software evolution and supporting the analysis of the impact of changes on system behavior. We present a motivating example and a preliminary evaluation based on a cyber-physical system controlling unmanned aerial vehicles. The results of our preliminary evaluation show that our approach can help to analyze changed behavior and thus contributes to understanding software evolution. 
\end{abstract}


\section{Introduction}

The full behavior of large-scale systems such as software-intensive and cyber-physical systems (CPS) typically only fully emerges at run time. It has been shown that intentional changes, such as adding new features or fixing bugs, often lead to unintentional changes with further side-effects~\cite{wagner2016relationship}. Even small changes may significantly affect the overall system behavior. 
In the field of software maintenance and evolution a variety of approaches have thus been introduced to analyze changes and to better understand evolution. For instance, many approaches analyze the change impact by comparing source code~\cite{fluri2007change,maletic2004supporting,baxter1998clone} or model-based representations~\cite{soto2006process,alanen2003difference}, often supported by information extracted from version control systems~\cite{dintzner2016fever,canfora2007identifying}. 
Due to limitations of static analyses for determining the change impact on run-time behavior, techniques have been developed to analyze and compare execution logs produced by systems at run time~\cite{ThanhoferPilisch2017, miranskyy2012using, reiss2001encoding}.

In this paper, we propose a new approach for determining the change impact by applying our existing constraint mining algorithm~\cite{Krismayer2019_CAiSE} to analyze the execution logs of multiple software versions. Our mining technique detects and analyzes recurring event patterns to extract constraints on event occurrence, timing, and event data (as well as combinations of these types). It presents the identified candidate constraints to users in a domain-specific language (DSL)~\cite{RabiserASEJ2018} and offers a range of filtering and ranking strategies~\cite{Krismayer2019_REFSQ}. 
The approach presented in this paper uses constraint mining to re-mine constraints after changes to the system and then compares the constraint sets. 
Differences are analyzed to highlight important changes in system behavior to the users, thereby lifting the level of abstraction from source code and execution logs to differences of constraints in a DSL. 
Specifically, changes in behavior become visible by constraints only found before or after the change, constraints with different values~(e.g., different intervals for allowed values of event data items), or constraints changed in other ways~(e.g., changed event names). Obviously, constraints may also remain identical in both sets.  

In the remainder of this paper we use a motivating example from the domain of unmanned aerial vehicles~(UAVs) and discuss how the behavior of a cyber-physical UAV control system can change due to maintenance and evolution. We briefly describe the constraint mining technique we build on and then introduce our evolution analysis approach. We further present a preliminary evaluation of our approach by applying it to the UAV control system. We conclude with a discussion of the status and an outlook on future research.

\section{Motivating Example}

Dronology~\cite{cleland2018dronology} is a CPS providing a research environment for managing, monitoring, and coordinating the flights of multiple UAVs, i.e., drones. The system provides features to assign missions and to simultaneously control multiple UAVs of different types. It can interact with real hardware~(the flying physical UAVs) and with a fully-fledged, high-fidelity Software-in-the-Loop simulator that enables experimentation with virtual UAVs. 
Both physical and virtual UAVs are controlled by a dedicated ground-control station that handles commands sent to, and messages received from the UAV.

For example, when a new UAV connects to Dronology, it sends a \emph{PHYSICAL\_UAV\_ACTIVATED} event with the current position of the drone. After activation each connected UAV once per second sends a \emph{state} event containing its position, speed, battery status, flight mode, etc.
Dronology allows to define routes consisting of multiple waypoints and assign them to one of the connected UAVs. When starting and finishing a route, and at each waypoint, the system creates the respective events~(\emph{PLAN\_ACTIVATED}, \emph{PLAN\_FINISHED}, \emph{WAYPOINT\_REACHED}).

As the system evolves, new features are introduced, bugs are fixed, and new functionality is added to accommodate additional application scenarios. This inevitably results in new or changed events, as well as different event data elements and data values attached to events.
Furthermore, changes to the UAV hardware, for example, adding new sensors, or the use of new UAV models with different physical characteristics~(e.g., hexcopter vs. quadcopter) can have a big impact on system behavior, which in turn affects the events and data collected at run time. 
Such differences are not always the result of changes in the system, but can also result from executing different tasks. For example, a new maximum flight altitude might be the result of different routes assigned to the UAVs.

\section{Background: Constraint Mining}

Our existing constraint mining approach~\cite{Krismayer2019_CAiSE} extracts the behavior of a software system by analyzing execution logs collected during system runs. These logs contain complex events consisting of an event type, a timestamp, and~(optionally) event data elements, such as sensor values or status messages. The mined constraints can then be monitored to detect unexpected behavior in future runs. Specifically, the mining approach extracts three types of constraints: 

\textbf{Temporal constraints} define the order of multiple events and their timing. An example from Dronology is ``\keyword{if event} PLAN\_ACTIVATED \keyword{occurs event} PLAN\_FINISHED \keyword{occurs within} 1\,min''.

\textbf{Value constraints} check the validity of one or multiple data elements attached to events of one particular type. An example from Dronology is the constraint ``\keyword{if event} WAYPOINT\_REACHED \keyword{occurs} data("location.z") $>=$ 10''. It checks that a drone is at least 10\,m above the ground at every waypoint.  

\textbf{Hybrid constraints} combine temporal and value constraints, i.e., they check the correct order and timing of multiple events as well as the validity of one or more values from event data items. An example of a hybrid constraint is ``\keyword{if event} PHYSICAL\_UAV\_ACTIVATED \keyword{occurs event} state \keyword{where} PHYSICAL\_UAV\_ACTIVATED.data("x") $==$ state.data("location.x") \keyword{occurs within} 1\,s''.

After the preparatory step of parsing execution logs the \textbf{mining approach} extracts these types of constraints in three main steps. It then filters, groups, and ranks the mined constraints in a fourth step before presenting them to the users:

\emph{Detecting event sequences.} 
The algorithm first detects event sequences, i.e., recurring patterns of events that usually appear together. One example of such a sequence in Dronology is that \emph{PLAN\_ACTIVATED} is followed by \emph{PLAN\_FINISHED}. 
The sequences detected in this step are temporal constraints.

\emph{Creating feature vectors.} 
The algorithm then creates feature vectors for each instance of the found sequences, i.e., groups of events in the input log matching the respective pattern. Each vector contains the values of all event data elements of the events of the sequence instance.
Event data elements that have the same value for all vectors are extracted as value constraints and hybrid constraints before being removed from the vectors.

\emph{Analyzing feature vectors.} 
Finally, the algorithm analyzes the feature vectors for each of the sequence types. Value and hybrid constraints are generated for event data elements that have the same value for the majority of the vectors, for intervals that contain the values for one event data element~(not including outliers), and for multiple event data elements with equal values for all vectors. 

\emph{Filtering, grouping, and ranking constraints}. 
The mined constraints are then filtered, i.e., an algorithm eliminates duplicate and highly similar constraints. The algorithm also removes constraints with low confidence, i.e., the rate of positive evaluations among all evaluations, and~(optionally) low support, i.e., the absolute number of positive evaluations~\cite{Lo2012}. The algorithm groups the remaining constraints, ranks them~(e.g., based on their confidence), and presents them to domain experts who select the ones to be used for system monitoring or other applications~\cite{Krismayer2019_REFSQ}.

\section{Evolution Analysis Approach}

\begin{figure}[t]
    \centering
    \includegraphics[width=1.0\columnwidth]{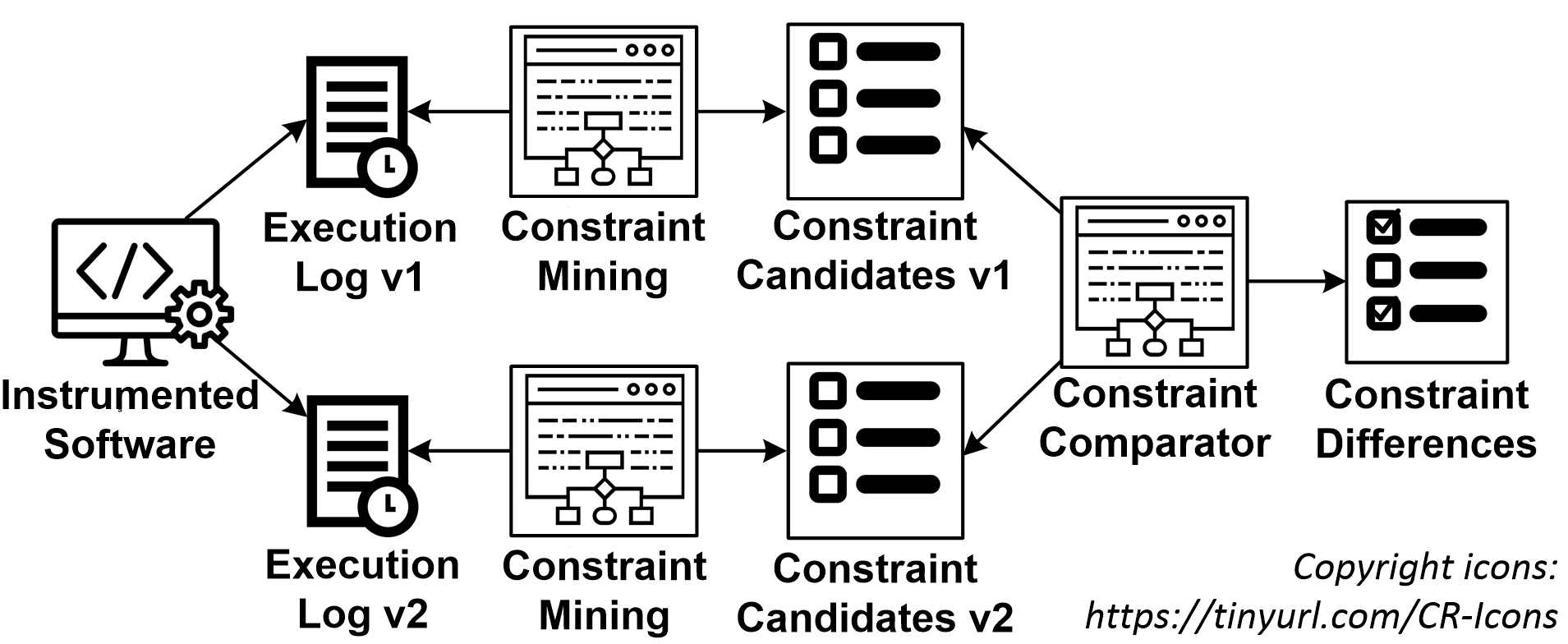}
    \caption{Evolution analysis approach.}
    \label{fig:approach}
\end{figure}

Our approach aims to support evolution by detecting and analyzing changes in the behavior of a system. These changes are reflected in disparities of the constraints mined from two different system versions. The approach shown in Figure~\ref{fig:approach} comprises steps for mining constraints and for comparing, analyzing, and presenting the mining results. 

\emph{Step 0: Extracting and preparing datasets.} The input required for the mining approach are system execution logs recorded from a system before and after the changes to be investigated. Each log represents a particular system version. It can be a simple log file as produced by most systems today or a more advanced execution log format. 

\emph{Step 1: Mining constraints for datasets.} 
The approach then uses constraint mining~\cite{Krismayer2019_CAiSE} to extract constraints in an existing DSL~\cite{RabiserASEJ2018} for both datasets. 
This step uses all the constraints produced by the constraint mining algorithm with a confidence of at least 90\%~(based on initial experiments) to filter irrelevant constraints and therefore does not require user input. 

\emph{Step 2: Comparing resulting constraint sets.} 
Our algorithm then compares the two resulting constraint sets to detect differences as illustrated in Algorithm~\ref{algo:compare_sets}. Specifically, it detects identical constraints, constraints with updated values, constraints with major changes (e.g., renamed events/event data elements or changes in the event sequence), and constraints only mined for one of the datasets. 
First, the algorithm detects identical constraints appearing in both sets and constraint pairs with the same structure~(same type, events, and event data element names) but different values~(e.g., different thresholds for data values or duration); cf. lines~\ref{line:identical} and \ref{line:differentVal} in Algorithm~\ref{algo:compare_sets}. 
These constraint pairs are then removed from the sets used for further analyses~(line~\ref{line:updateA}--\ref{line:updateB}).

The third group of constraint pairs are those with major differences, e.g., differences in the event sequence or renamed events and event data elements. For this type of matching we calculate a similarity score between all remaining constraints from the two lists and only match two constraints, if the similarity between them is above 0.5. If one constraint could be matched to multiple constraints in the other list, the one with the highest similarity is chosen~(cf. line~\ref{line:different} in Algorithm~\ref{algo:compare_sets}). 
The similarity is calculated based on the type of the event that triggers the constraint evaluation, the overall sequence alignment, the duration values, and the types and values of the event data elements of two constraints. 

The remaining constraints are then representing behavior that is only found for one of the two datasets. In the case of logs recorded from different versions of the same software system these constraints thus indicate behavior only observed before or after the update, i.e., missing and new constraints~(cf. lines~\ref{line:missing}--\ref{line:new} in Algorithm~\ref{algo:compare_sets}).

\begin{algorithm}[t]
    \caption{Compare constraint sets.}
    \label{algo:compare_sets}
    \begin{algorithmic}[1]
    
    \Function{detectDifferences}{$\mathit{A}, \mathit{B}$}
        
        \State $\mathit{Identical}\gets \{(\mathit{a},\mathit{b}) \mid (a,b)\in A\times B \land \mathit{a}=\mathit{b} \}$ \label{line:identical}
        
        \State \begin{varwidth}[t]{\linewidth}$\mathit{DifferentVal}\gets \{(a,b) \mid (a,b)\in A\times B\, \land$ \par 
        \hspace\algorithmicindent $\mathit{structure(a)}=\mathit{structure(b)}\, \land$ \par 
        \hspace\algorithmicindent $\mathit{values(a)}\neq\mathit{values(b)} \}$
        \end{varwidth} \label{line:differentVal}
        
        \State $A\gets A\setminus \{a \mid (\exists b \mid (a,b)\in \mathit{Identical} \cup \mathit{DifferentVal)}\}$\label{line:updateA}
        
        \State $B\gets B\setminus \{b \mid (\exists a \mid (a,b)\in \mathit{Identical} \cup  \mathit{DifferentVal)}\}$\label{line:updateB}
        
        \State \begin{varwidth}[t]{\linewidth} 
        $\mathit{Different}\gets \{(a,b) \mid (a,b)\in A\times B\, \land$ \par 
    	\hspace\algorithmicindent $\mathit{similarity(a,b)} > 0.5 \land (\nexists c \mid c\in B\, \land $ \par 
    	\hspace\algorithmicindent $similarity(a,c) > similarity(a,b)) \}$
    	\end{varwidth} \label{line:different}
        
        \State $\mathit{Missing}\gets A\setminus \{a \mid (\exists b \mid (a,b)\in \mathit{Different}) \}$ \label{line:missing}
        
        \State $\mathit{New}\gets B\setminus \{b \mid (\exists a \mid (a,b)\in \mathit{Different}) \}$ \label{line:new}
        
        \State \Return $\mathit{Identical}, \mathit{DifferentVal}, \mathit{Different},$ 
        \Statex \hskip \algorithmicindent \hskip \algorithmicindent $\mathit{Missing}, \mathit{New}$
        
    \EndFunction
    \end{algorithmic}
\end{algorithm}

\emph{Step 3: Analyzing, ranking and presenting the results.} 
In the final step we present the lists of matched constraints to the user. To make the results easier to evaluate we present the constraints as separate lists (identical, different values, different, missing, and new) and additionally show the evaluation results~(positive and total number of evaluations and confidence) on both datasets for each constraint. 

We further rank the lists based on different criteria. The list of identical constraints is ranked based on the total number of positive evaluations. We sort the constraint pairs with different values and the constraints with major differences based on the similarity of the pairs~(higher similarity is ranked higher). Finally, the two lists with unmatched constraints, i.e., missing and new constraints, are ranked such that constraints with high confidence for the dataset they are mined from and low confidence for the other dataset are ranked highest.

\section{Evaluation}
\label{sec:eval}

The goal of our preliminary evaluation is to demonstrate the basic feasibility and usefulness of our approach. We recorded two different runs from the Dronology simulator with five UAVs flying ten distinct routes~(each consisting of two to five waypoints) at a dedicated flying field. Each route was randomly assigned twice to one of the drones. For the two runs, two different sets of routes were randomly generated. After the assigned routes were completed the drones returned to their starting location. 

We used these two datasets to investigate two research questions in two experiments:

\emph{Baseline Comparison (RQ1): Does our approach discover expected minor differences in logs of two executions of the same software version?} In the first experiment, we directly used the two execution logs as input and compared the constraints mined from them. Since both datasets contained logs from the same system with no actual changes made to the system, we expected to find only minor differences in the mined constraints resulting from differences in values, e.g., different flying altitude and routes of drones. Experiment 1 was thus aimed to show the basic feasibility of our approach. 

\emph{Seeded Changes Comparison (RQ2): Does our approach discover differences in logs of two executions of different software versions simulated via seeded changes in one dataset?} In the second experiment, we introduced a number of changes to the events in the second log. These changes simulate different types of modifications made to the source code of the system (e.g., renaming events in the log statements) and in the behavior of the UAVs (e.g., different maximum flight altitude). Specifically, we expected our approach to discover identical constraints, new constraints and missing constraints, as well as constraints with major differences and with differences only in data values. Experiment 2 thus aimed to show the feasibility and usefulness of our approach by demonstrating that it can discover different types of changes by analyzing differences in two logs of executions of a changed software system.

\subsection{Experiment 1: Baseline Comparison}

Comparing the two datasets revealed that 44 of the mined constraints were identical while 36 exhibited differences in values. Table~\ref{tab:examples_1} provides examples of one identically mined constraint~(\#1) and two constraints with changed values: \#2~with \emph{MAX} as 14.25 for the first dataset and 14.089 for the second dataset and \#3~with \emph{T} being 1\,s for the first dataset and 2\,s for the second dataset. 
One temporal constraint was detected as changed, because in the sequence the \emph{monitoring} and \emph{state} event were swapped. These two events are both sent approximately once per second and it is impossible to tell which will arrive first. Such minor differences could be expected when comparing two different executions of the same software with slightly different tasks.

None of the mined constraints appeared only in the first dataset, while ten constraints could only be found in the second dataset. As it turns out all of these constraints were not mined for the first dataset, because their confidence was below 90\%. An example for such a constraint is \#4 in Table~\ref{tab:examples_1}, which has a confidence of 90.5\% for the second dataset, but only 81.9\% for the first one. Such different confidence levels can be expected due to the different scenarios and resulting actual behavior during the two executions, in this case the different flying times~(in \emph{GUIDED} mode) vs. idle times. 

These observations show that our approach is capable of discovering (expected)~differences in two logs of executions of the same software.

\begin{table}[t]
\centering
\caption{Examples of constraints mined in experiment 1.}
\label{tab:examples_1}
\begin{tabular}{ll}
\toprule
Nr & Constraint \\ \midrule
1 & \keyword{if event} monitoring \keyword{occurs} \\
& data("voltage") >= 12.094 \keyword{and} data("voltage") <= 12.587 \\ 
\addlinespace[0.25em]
2 & \keyword{if event} monitoring \keyword{occurs} \\
& data("groundspeed") >= 0.0 \keyword{and} data("groundspeed")  <= \emph{MAX} \\
\addlinespace[0.25em]
3 &  if event "PLAN\_COMPLETE" occurs \\
& event "state" where data("status") == "ACTIVE" occurs within \emph{T} \\
\addlinespace[0.25em]
4 & trigger = if event "state" occurs data("mode") == "GUIDED" \\
\bottomrule
\end{tabular}
\end{table}

\subsection{Experiment 2: Seeded Changes Comparison}

For this experiment we seeded six changes to the second execution log~(cf. Table~\ref{tab:changes}) and compared the mined constraints. Our algorithm found 35 identical constraints, 33 constraints with different values, eleven constraints with major differences, two constraints only mined for the first dataset, and 13 constraints only mined for the second dataset.

\begin{table}[t]
\centering
\caption{Changes applied to the second dataset.} 
\label{tab:changes}
\begin{tabular}{cll}
\toprule
Nr & Seeded change & Observed differences  \\ \midrule

\multirow{2}{*}{1} & added new event data element ``rnd''  & \multirow{2}{*}{one new constraint} \\
& with random values to state events  \\ 
\addlinespace[0.25em]

\multirow{2}{*}{2} & increase value of data element  & one constraint with \\
& ``location.z'' in state events by 5\% & different values \\ 
\addlinespace[0.25em]

\multirow{2}{*}{3} & increase value of data element  & one constraint with \\ 
& ``altitude'' in monitoring events by 5\% & different values \\ 
\addlinespace[0.25em]

4 & 
\begin{tabular}[c]{@{}l@{}} WAYPOINT\_REACHED event\\ renamed to waypoint \end{tabular} & 
\begin{tabular}[c]{@{}l@{}} ten changed constraints,\\ one missing constraint, \\ three new constraints \end{tabular} \\ 
\addlinespace[0.25em]

\multirow{2}{*}{5} & delete data element ``n\_satellites'' from & \multirow{2}{*}{one missing constraint}\\
& monitoring events \\ 
\addlinespace[0.25em]

\multirow{2}{*}{6} & rename data element ``start-time'' in & \multirow{2}{*}{one changed constraint}\\
& PLAN\_COMPLETE events to ``begin'' \\
\bottomrule
\end{tabular}
\end{table}

Our first change was to add an additional event data element to the events of type \emph{state}. 
This change is equivalent to reporting an additional sensor value or a value related to an internal state.
The data element was correctly extracted as a new value constraint, cf. constraint \#1 in Table~\ref{tab:examples_2}.

The second and third change applied to the dataset affected the altitude values of both \emph{state} and \emph{monitoring} events, i.e., simulating that the drones fly higher. 
A change in the source code that would lead to the same differences could be a bugfix for the calculation of the altitude from the sensor values or updated geographic location data.
The different altitude could also result from a different maximum altitude allowed for a new drone type or a different regulation for the flying field. 
Both changes were detected in the form of matching constraints with different values, cf. constraints \#2 and \#3 in Table~\ref{tab:examples_2}. For \#2 \emph{MAX} is 28.3 for the first dataset and 31.049 for the second dataset; for \#3 the intervals are [-0.069; 28.285] for the first dataset and [-0.072; 31.035] for the second dataset.

Changing the type \emph{WAYPOINT\_REACHED} to \emph{waypoint}---the fourth change from Table~\ref{tab:changes}---simulates, e.g., an update in the logging, a change in the process, or a refactoring. 
As a result ten constraints were detected as constraints with major differences~(one example is \#4 in Table~\ref{tab:examples_2}, which uses \emph{waypoint} instead of \emph{WAYPOINT\_REACHED}), one constraint only found for the first dataset, and three constraints only found for the second dataset. 

The fifth change, i.e., deleting the event data element \emph{n\_satellites} from all events of type \emph{monitoring}, could result from removing a data element in the source code or from a sensor or a measurement that is no longer available. This change lead to a missing constraint, i.e., constraint \#5 from Table~\ref{tab:examples_2} was only mined for the first dataset. 

Renaming the event data element \emph{start-time} to \emph{begin}, the last change listed in Table~\ref{tab:changes}, could result from renaming a field in the source code or changing the name in the logger. 
This change resulted in one constraint (\#6 in Table~\ref{tab:examples_2}) with major differences. Not only the name of the event data element has changed, but also the upper and lower bound for its values---resulting from the different execution times of the two runs.

These observations confirm that our approach can discover differences in two execution logs of changed software, demonstrated via seeded changes. 

\begin{table}[t]
\centering
\caption{Examples of constraints mined in Experiment 2.}
\label{tab:examples_2}
\begin{tabular}{lll}
\toprule
Nr & Constraint & Change type \\ \midrule

\multirow{2}{*}{1} & \keyword{if event} state \keyword{occurs} & \multirow{2}{*}{new} \\
& data("rnd") >= 0.01 \keyword{and} data("rnd") <= 0.989 \\ 
\addlinespace[0.25em]

\multirow{2}{*}{2} & \keyword{if event} monitoring \keyword{occurs} & \multirow{2}{*}{different value} \\
& data("alt") >= 0.0 AND data("alt") <= \emph{MAX} \\
\addlinespace[0.25em]

\multirow{3}{*}{3} & \keyword{if event} state \keyword{occurs} & \multirow{3}{*}{different value} \\
& data("location/z") >= \emph{MIN} \\
& \keyword{and} data("location/z") <= \emph{MAX} \\
\addlinespace[0.25em]

\multirow{3}{*}{4} & \keyword{if event} waypoint \keyword{occurs} & \multirow{3}{*}{different} \\
& state \keyword{where} data("status") == "ACTIVE" \\ 
& \keyword{occurs within} 2\,s \\
\addlinespace[0.25em]

\multirow{2}{*}{5} & \keyword{if event} monitoring \keyword{occurs} & \multirow{2}{*}{missing} \\
& data("n\_satellites") == 10 \\
\addlinespace[0.25em]

\multirow{3}{*}{6} & \keyword{if event} PLAN\_COMPLETE \keyword{occurs} & \multirow{3}{*}{different} \\
& data("start-time") >= 1559562125924 \\
& \keyword{and} data("start-time") <= 1559562873612 \\

\bottomrule
\end{tabular}
\end{table}

\subsection{Threats to Validity}

One threat to validity is the use of a single application scenario with seeded changes. However, even our preliminary evaluation based on two experiments  allowed us to assess all possible types of differences in two sets of mined constraints and the results indicate the basic feasibility of our approach. Also, the seeded changes affect the events in the same way as specific changes to the source code would. 
Another threat to validity is that our detection relies on lists of constraints mined for the input datasets. These lists could be incomplete and could contain irrelevant constraints, i.e., in the context of this paper we assume the mining is correct. 
While the sets of mined constraints may not be perfect, we still argue that they can give a good overview of the relevant behavior of the software system and can be used to detect many changes between two system versions. Domain experts have confirmed the usefulness---e.g., for actually monitoring their systems---of the mined constraints in past experiments~\cite{Krismayer2019_CAiSE}.

\balance

\section{Status and Next Steps}
\label{sec:status_and_next_steps}

We presented an approach to detect changes in the behavior of a software system based on automatically mined constraints. Our preliminary evaluation showed that the approach correctly matches mined constraints and finds diverse kinds of differences, thereby indicating basic feasibility and usefulness. 
However, there are several \emph{opportunities} to improve our current implementation:

Our approach currently works without any \emph{user involvement}. In the future we plan to experiment with a feature allowing domain experts to provide feedback on the usefulness of the initially mined constraints. Utilizing their ratings would allow the algorithm to focus on the most relevant differences. 

Usually some of the constraints mined for a system need to be slightly modified before being used for run-time monitoring. Such modifications include adapting thresholds or intervals allowed for certain data values or the time allowed for a specific sequence of events. 
For example, the mining algorithm extracts the maximum duration of a UAV executing one route as four minutes. However, in practice this duration might be allowed to be longer in some situations. This information, i.e., \emph{adapted constraints}, can also be utilized in the comparison. 

Furthermore, we plan to extend the configuration options of the approach including \emph{different ranking algorithms and search capabilities} to further support the users in analyzing and selecting relevant constraints. 

While sorting the lists can ease the task for the users, this still requires some usability improvements for large numbers of constraints. 
We thus also plan to implement \emph{tool support} for presenting the output of our comparison algorithm.  

To further evaluate the approach we plan to \emph{apply the algorithm to other domains and systems}. Specifically, we will experiment with different types of systems producing other kinds of execution logs. For example, while Dronology produces many events in a short time, other systems produce fewer events over longer periods of time, which might have an effect on the mined constraints and the comparison results.

We also plan to use our approach to \emph{detect differences} between constraints mined \emph{for different scenarios} and applications of a software system. This information might help to understand how the behavior of the system and certain requirements change depending on the scenario.

Finally, we plan to \emph{improve the constraint matching} based on previous feedback, e.g., using machine learning techniques.

\section*{Acknowledgements}
The financial support by the Austrian Federal Ministry for Digital and Economic Affairs, the National Foundation for Research, Technology and Development, Primetals Technologies, and the Austrian Science Fund (FWF) under Grant No. J3998-N31 is gratefully acknowledged.

\balance
\bibliographystyle{plain}
\bibliography{literature.bib}

\end{document}